\documentclass[epjCONF,columns]{svjour} 
\usepackage{graphics}
\usepackage[varg]{txfonts} 
\usepackage[latin1]{inputenc}
\session-title{Hadron Collider Physics Symphosium 2011, Paris (France)}
\begin{document}
\title{$\rm D^{+}_{s}$ production at central rapidity in pp collisions at 7 TeV
with the ALICE experiment}
\author{Gian Michele Innocenti\inst{1}\fnmsep\thanks{\email{ginnocen@cern.ch}} for the ALICE Collaboration}
\institute{Department of Experimental Physics, University $\&$ INFN, Turin, ITALY}
\abstract{
We present the preliminary $p_{\rm t}$ differential cross section in pp collisions of the $\rm D^{+}_{s}$ meson
measured in the mid-rapidity region of ALICE through the $\rm D^{+}_{s} \rightarrow \rm K^{+}K^{-}\pi^{+}$ 
decay channel with an integrated luminosity of 4.8 nb$^{-1}$. 
The ratios between all the D meson preliminary $p_{\rm t}$ differential cross sections
measured in the ALICE experiment ($\rm D^{+}_{s}$, $\rm D^{0}$, $\rm D^{+}$, $\rm D^{*+}$) are also presented and 
compared with the results of other experiments.}
%
\maketitle
\section{Introduction}
\label{intro}
The measurement of the charm production cross section in pp collisions 
is a fundamental test for the perturbative QCD calculations in the new energy regime of the LHC. 
In particular, the measurement of the $\rm D^{+}_{s}$ production allows to study
the fraction and the $p_{\rm t}$ distribution of the charmed-strange mesons.
Results in pp collisions also provide a crucial reference for Pb-Pb studies
in which heavy quarks are expected to be important probes for the properties of the medium \cite{piombo}.
In these proceedings, we report on the measurement of the production cross section of  
prompt $\rm D^{+}_{s}$ mesons in pp collisions reconstructed in the transverse momentum 
range 2 $<$ $p_{\rm t}$ $<$ 12 GeV/$c$ at central rapidity ($|\rm y|$ $<$ 0.5) with the ALICE 
detector, using data collected in 2010.
\section{Detector layout and data sample}
\label{sec:1}
The D mesons are reconstructed in the central rapidity region using the 
tracking detectors and particle identification systems of the ALICE central barrel 
which are placed in a large solenoid magnet, with a
field B = 0.5 T, and cover the pseudo-rapidity region -0.9 $ < \eta < $ 0.9.
The central barrel detectors allow to track charged particles down 
to low transverse momenta ($\approx$ 100 MeV/c) and provide charged hadron and electron 
identification together with an accurate measurement of the positions of the primary
and secondary vertices~\cite{alice}.
In this section, a short description of the detectors utilized in
these analyses will be given (See \cite{alice} for further details).
The closest detector to the beam axis is the Inner Tracking System (ITS) \cite{its} which
is composed of six cylindrical layers of silicon detectors.
The two innermost layers (at radii of $\approx$ 4 and 7 cm) are made of pixel detectors (SPD), 
the two intermediate layers (radii $\approx$ 15 and 24 cm) are equipped with drift detectors, 
while strip detectors are used for the two outermost layers (radii $\approx$ 39 and 44 cm).
The ITS allows the detection of secondary vertices 
originating from open charm decays with a resolution on the impact parameter
better than 50 $\mu$m for tracks with $p_{\rm t}$ $>$ 1.3 GeV/$c$~\cite{paperoD}.
The Time Projection Chamber (TPC) \cite{tpc} is the main tracking detector that provides track 
reconstruction and particle identification via the measurement of the specific energy 
deposit dE$/$dx.
The Time-of-Flight (TOF) detector is used for pion, kaon and proton identification 
on the basis of their time of flight and it provides kaon/pion separation up to a 
momentum of about 1.5 GeV/$c$. All the three detectors have full azimuthal coverage.
The data sample used for this analysis consists of $\approx$ 300 million minimum-bias (MB)
events collected during the 2010 LHC run with pp collisions at $\sqrt{s}$ = 7 TeV
which correspond to an integrated luminosity $\rm L_{int}$ = 4.8 nb$^{-1}$.
The minimum-bias trigger was based on the SPD and VZERO
detectors. The latter is made of two scintillator hodoscopes positioned in
the forward and backward regions of the experiment \cite{alice}.
The cross section of pp collisions passing the the MB condition used 
for the $\rm D^{+}_{s}$ production cross section normalization, was derived from a measurement 
of the cross section of collisions that give signals in both sides of the VZERO detector
using a van der Meer scan \cite{scan}.

\section{D meson production measurements in ALICE}
\label{sec:2}
\begin{figure*}
\hspace{2cm}
\resizebox{.65\textwidth}{!}{%
  \includegraphics{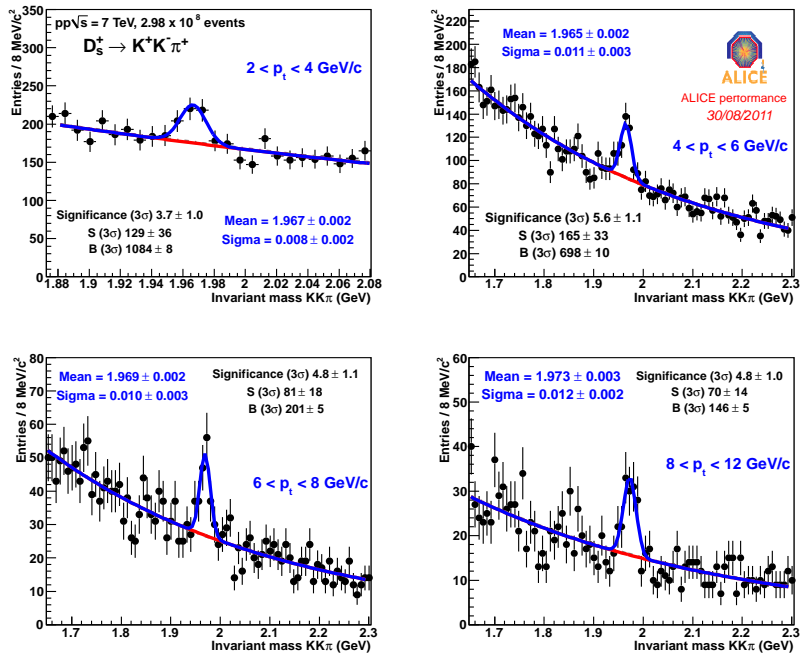} 
  }
\caption{ Invariant mass distributions of $\rm D_{s}^{+}$ candidates in 4 $p_{\rm t}$ bins in the transverse momentum range 
    2 $<$ $p_{\rm t}$ $<$ 12 GeV/$c$ obtained from the analysis of $\approx$ 300 million minimum bias events.}
\label{invmass}       
\end{figure*}

\begin{figure}
\resizebox{0.75\columnwidth}{!}{%
  \includegraphics{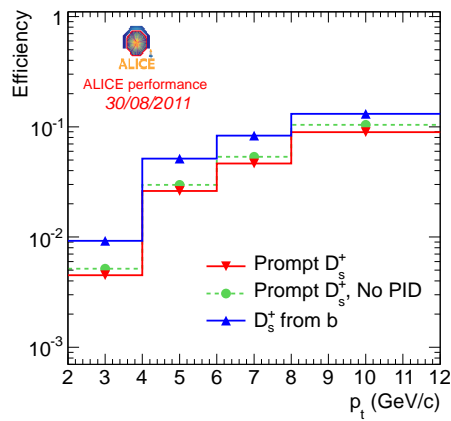} 
  }
\caption{Efficiency for the $\rm D_{s}^{+}$ meson as a function of $p_{\rm t}$ for prompt D mesons
    (with and without PID selection) and D mesons from B feed-down.}
\label{eff}       
\end{figure}

The measurement of $\rm D^{+}_{s}$ meson production was performed by reconstructing the 
$\rm D^{+}_{s} \rightarrow \rm K^-K^+\pi^+$ decay channel (with a branching ratio, BR, of 5.49 $\pm$ 0.27 $\%$ \cite{pdg})
with its charge conjugate. 
In this analysis the resonant channel with a $\rm \phi$ in the intermediate state, $\rm D_{s}^{+} \rightarrow \phi \pi^{+} \rightarrow K^{+} K^{-} \pi^{+}$, was considered. 
The analysis strategy for the extraction of the $\rm D^{+}_{s}$ signals 
out of the large combinatorial background is based on the reconstruction and selection of 
secondary vertex topologies with significant separation from the primary vertex.
At first, single tracks were selected with respect to to their momentum and pseudorapidity 
($p_{\rm t}$ $>$ 0.4 GeV/$c$ and $|\eta|$ $<$ 0.8) and according to quality cuts.
The D meson candidates were then built starting from track combinations with proper charges
and selected using topological cuts.
The candidate triplets were selected according to the sum of the distances of the decay tracks 
to the reconstructed decay vertex, the decay length and the cosine of the pointing angle, 
that is the angle between the reconstructed D meson momentum and the line which connects the 
primary and the secondary vertex.
In addition, since the final state considered occurs via resonant channel with a $\rm \phi$
in the intermediate state, the decay $\rm D_{s}^{+} \rightarrow \phi \pi^{+} \rightarrow K^{+} K^{-} \pi^{+}$
was selected by requiring that one of the two pairs of opposite-signed tracks has 
invariant mass compatible with the $\rm \phi$ mass.
A particle identification (PID) strategy, that uses the specific energy deposit from the TPC and 
the time-of-flight from the TOF, has also been adopted to provide a further reduction of the 
background while preserving most of the $\rm D_{s}^{+}$ meson signal.
The invariant mass distributions of the candidates were then fit with a function that 
consists of a Gaussian term describing the signal and an exponential
term for the background. In Fig. \ref{invmass} the invariant mass distributions of $\rm D_{s}^{+}$ 
candidates in 4 $p_{\rm t}$ bins in the transverse momentum range 2 $<$ $p_{\rm t}$ $<$ 12 GeV/$c$ 
are shown. These results are obtained from the analysis of $\approx$ 300 million pp minimum bias events.
The raw signal yields obtained from the invariant mass analysis in each $p_{\rm t}$ bin 
have been corrected for acceptance (which also considers the rapidity range of the cross 
section measurement $|\rm y|$ $<$ 0.5) and selection efficiency of prompt D mesons using 
Monte Carlo simulations based on the PYTHIA 6.4.21 event generator~\cite{pythia} 
with Perugia-0 tuning~\cite{perugia}.
The contribution of $\rm D_{s}^{+}$ mesons coming from B meson decays (B feed-down) 
has been evaluated using the Monte Carlo efficiency for feed-down D mesons and the 
FONLL pQCD calculation which well describes
bottom production at Tevatron~\cite{teva} and at the LHC~\cite{lhc1,lhc2}.
The fraction of D meson from b quark decays has been estimated as 10 - 15 $\%$ depending on the 
$p_{\rm t}$ of the D meson. In Fig. \ref{eff} the efficiency for the
$\rm D_{s}^{+}$ meson with $|\rm y|$ $<$ 0.5 as a function of the transverse momentum $p_{\rm t}$
for prompt $\rm D_{s}^{+}$ mesons and $\rm D_{s}^{+}$ mesons from B feed-down decays is shown. 
D mesons from b quark decays present a larger efficiency since they decay further from the
primary vertex because of the large B meson lifetime (c$\tau$ $\approx$ 500 $\mu$m \cite{pdg}).

\section{Results}
\label{sec:3}
\begin{figure}
\resizebox{0.75\columnwidth}{!}{%
  \includegraphics{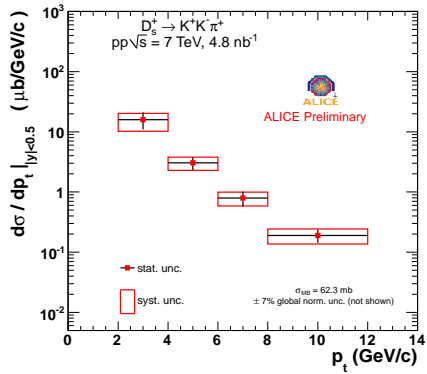} 
  }
\caption{Preliminary $\rm D_{s}^{+}$ $p_{\rm t}$ differential cross section measured 
    with an integrated luminosity of 4.8 nb$^{-1}$ ($\approx$ 300 million minimum-bias events)
    in the $p_{\rm t}$ range 2-12 GeV/c.}
\label{Dscross}       
\end{figure}

\begin{figure}
\resizebox{0.75\columnwidth}{!}{%
  \includegraphics{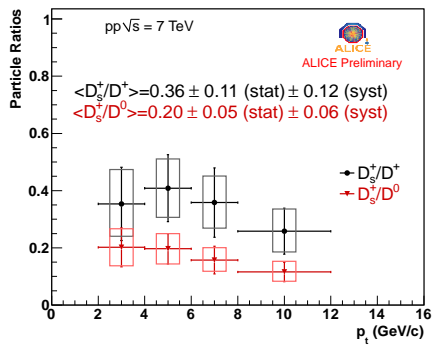} 
  }
\caption{Ratios between the preliminary D meson $p_{\rm t}$ differential cross sections $\rm D_{s}^{+}$/$\rm D^{0}$
and $\rm D_{s}^{+}$/$\rm D^{+}$ in the $p_{\rm t}$ range 2 $<$ $p_{\rm t}$ $<$ 12 GeV/$c$.
 Both statistical and systematic uncertainties are shown. The $p_{\rm t}$ integrated values are also
 reported in the legend.}
\label{ratio}
\end{figure}

\begin{figure}
\resizebox{0.85\columnwidth}{!}{%
  \includegraphics{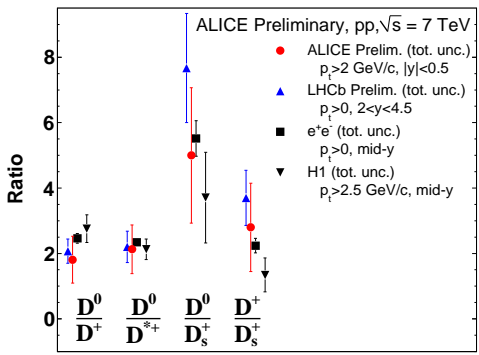} 
  }
\caption{Ratios between the preliminary D meson cross sections 
    integrated in the $p_{\rm t}$ range 2 $<$ $p_{\rm t}$ $<$ 12 GeV/$c$ measured by ALICE compared with 
    the results from other experiments~\cite{lhcb,lep1,h1}. Total uncertainties are shown.}
\label{ratioAll}
\end{figure}

In Fig. \ref{Dscross} the preliminary $p_{\rm t}$ differential cross sections for prompt 
$\rm D_{s}^{+}$ mesons is shown in 4 $p_{\rm t}$ bins in the transverse momentum range
2 $<$ $p_{\rm t}$ $<$ 12 GeV/$c$. The error bars represent the statistical uncertainties, 
while the systematic uncertainties are shown as boxes around the data points. 
A prediction for production cross section of this meson is still not available and thus a 
direct comparison with the theoretical calculations is not yet feasible. 
Figure \ref{ratio} shows the preliminary results for the $p_{\rm t}$ differential 
cross section ratios $\rm D_{s}^{+}$/$\rm D^{0}$ and $\rm D_{s}^{+}$/$\rm D^{+}$ 
in the $p_{\rm t}$ range 2 $<$ $p_{\rm t}$ $<$ 12 GeV/$c$ with statistical and systematic
uncertainties together with the $p_{\rm t}$-integrated values. Due to the large uncertainties 
of the measurements, a final statement on the $p_{\rm t}$ dependence of these ratios is 
not yet possible.
Finally, in Fig. \ref{ratioAll} the ratios between the preliminary D meson cross sections integrated 
in the $p_{\rm t}$ range 2 $<$ $p_{\rm t}$ $<$ 12 GeV/$c$ ($\rm D^{0}$/$\rm D^{+}$, $\rm D^{0}$/$\rm D^{*+}$, 
$\rm D^{0}$/$\rm D_{s}^{+}$ and $\rm D^{+}$/$\rm D_{s}^{+}$ ) measured by ALICE
are compared with the results of other experiments, namely LHCb~\cite{lhcb}, 
$\rm e^{+}e^{-}$ data~\cite{lep1}, H1~\cite{h1}: 
ALICE results are compatible with the other measurements within the uncertainties.

\section{Conclusions}
\label{sec:4}
In these proceedings, we presented the preliminary measurements by the ALICE Collaboration
of the production cross sections of prompt $\rm D_{s}^{+}$ mesons in pp collisions at $\sqrt{s}$ = 7 TeV 
in the range 2 $<$ $p_{\rm t}$ $<$ 12 GeV/$c$ and in the central rapidity region. 
$\rm D_{s}^{+}$ mesons were reconstructed via their hadronic decay channel
$\rm D_{s}^{+} \rightarrow K^{+} K^{-} \pi^{+}$. We also reported on the ratios between the 
preliminary $p_{\rm t}$ differential cross section of D mesons compared with the results of
other experiments: ALICE results are compatible with the other measurements within the uncertainties.
%
%

\end{document}